\documentclass[aps,twocolumn,superscriptaddress,showpacs]{revtex4}
\usepackage{graphicx}
\begin{document}
\draft
\title{\bf  
The Spin-1 Heisenberg Antiferromagnet: New Results from Series Expansions}
\author{
J. Oitmaa and C.J. Hamer}
\affiliation{School of Physics, The University of New South Wales,
  Sydney, NSW 2052, Australia}
\date{\today}
\begin{abstract}
We calculate ground state properties (energy, magnetization,
susceptibility) and one-particle spectra for the $S = 1$ Heisenberg
antiferromagnet with easy-axis or easy-plane single site anisotropy, on
the square lattice. Series expansions are used, in each of three phases
of the system, to obtain systematic and accurate results. The location
of the quantum phase transition in the easy-plane sector is determined.
The results are compared with spin-wave theory.
\end{abstract}
\pacs{PACS Indices: 05.30.-d,75.10.-b,75.10.Jm,75.30.Ds,75.30.Kz \\
\\  \\
(Submitted to  Phys. Rev. B) }
\maketitle
\newpage

\section{Introduction}
\label{sec1}

Magnetic materials with S = 1 ions have been of interest for many years. The `classic' 2-dimensional 
Heisenberg antiferromagnet K$_2$NiF$_4$ was studied in the 1970s \cite{birgeneau1970}. In the 1980s a number of 
(weakly coupled) linear chain systems were investigated, including CsNiCl$_3$ \cite{steiner1987}, which has a 
weak axial anisotropy, CsFeBr$_3$ \cite{dorner1988}, which has strong planar anisotropy, and the complex materials 
NENP (Ni(C$_2$H$_8$N$_2$)$_2$NO$_2$(ClO$_4$)) \cite{renard1988} 
and NENC (Ni(C$_2$H$_8$N$_2$)$_2$Ni(CN$_4$)) \cite{orendac1995}, which have weak and strong planar anisotropy, 
respectively. The spin gaps observed in the weakly anisotropic materials \cite{steiner1987,renard1988} are believed 
to be 
examples of the behaviour predicted by Haldane \cite{haldane1983}.
 More recent work includes molecular oxygen adsorbed on 
graphite \cite{murakami1996}, the bilayer material Ba$_3$Mn$_2$O$_8$
\cite{uchida2002,stone2008}, a new spin gapped material NiGa$_2$S$_4$ \cite{nakatsuji2005}, which, it has been 
argued \cite{tsunetsugu2006,bhattacharjee2006}, 
may be a `spin nematic' \cite{chandra1991}, and a system of spin-1 bosonic atoms in
an optical lattice \cite{greiner2002}.

We have previously used series expansion methods to study a wide range of spin-1/2 quantum antiferromagnets 
\cite{oitmaa2006}. Quantum fluctuations will be reduced when S = 1, but new physical features are possible. Besides 
the gapped Haldane phase in 1-dimensional systems \cite{haldane1983}, additional terms, such as biquadratic exchange 
and/or single site anisotropy, which are absent in spin-1/2 systems, can lead to quadrupolar/nematic phases 
with long range order but no magnetic moment, novel quantum phase transitions \cite{sachdev1999}, and richer excitation spectra. 
We explore some of these issues in the present work, within the context of the Hamiltonian

\begin{equation}
H = J \sum_{<ij>} {\bf S_i \cdot S_j} - D \sum_i (S_i^z)^2
\label{eq1}
\end{equation}
which describes a Heisenberg antiferromagnet ($J > 0$) with isotropic exchange and a single ion anisotropy which 
gives rise to an easy axis ($D > 0$) or easy plane ($D < 0$).

A significant single ion anisotropy is believed to be present in many of the 1-dimensional
 materials referred to above, 
and has been included in analyses of the experimental data. Consequently there has been much theoretical work 
devoted to the Hamiltonian (\ref{eq1})  on a linear chain \cite{golinelli1992,papanicolaou1990,chen2003}. For higher 
dimensions various approaches 
have been used, including mean-field type theories \cite{devlin1971,khajehpour1975}, spin wave approximations 
\cite{rastelli1974,lindgard1975} and a coupled 
cluster calculation \cite{wong1994}. We will compare our results with each of these, where possible.

\begin{figure}
 \includegraphics[width=1.0\linewidth]{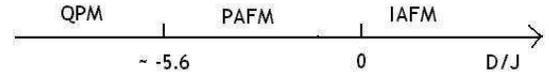}
\caption{ 
The phase diagram of the spin-1 J-D model on the square lattice at zero
temperature, showing the Ising antiferromagnetic phase (IAFM), planar antiferromagnetic
phase (PAFM), and quantum paramagnetic phase (QPM).
}
 \label{fig1}
\end{figure}

The present work deals with the 2-dimensional square (SQ) lattice. For D = 0 the spin-1 system has N{\' e}el 
order at zero temperature \cite{neves1986}, with quantum fluctuations reducing the staggered magnetization by some 20\% 
\cite{singh1990}. When $D > 0$ the system will order along $z$, the `easy axis'. The order parameter will be `Ising-like' 
and long-range order will persist at finite temperature, up to a critical line $T_c(D)$ with Ising (n=1) 
exponents. For $D < 0$, on the other hand, the z-axis is a `hard' direction and the spins will order 
antiferromagnetically along some direction in the x-y plane (at least for small $|D|$). The order parameter has n=2 
components and the continuous symmetry will be spontaneously broken.  Long-range magnetic order will not persist 
to finite temperature (the Mermin-Wagner theorem) although one may expect a Kosterlitz-Thouless transition. For 
large negative D the physics will be quite different. In the limit $D \rightarrow - \infty$ the ground state 
will be a simple product state with $S^z = 0$ at all sites. This is a quadrupole state with no magnetic order. 
Thus we expect a quantum phase transition at some $D = D_c$, which we will locate using our series approach. 
The phase diagram is illustrated in Figure \ref{fig1}.

It is also of interest to study the elementary excitations above the ground state. For D = 0 these are magnons, 
with Goldstone modes at ${\bf k} = (0,0)$ and $(\pi,\pi)$. A gap will open in the spectrum for small positive D, 
which will be proportional to $\sqrt{D}$, according to spin-wave theory. On the other hand for large $|D|$ the 
picture will be quite different. For large negative D the excitations will consist of isolated $S^z = \pm 1$ 
spins, which have been termed excitons and anti-excitons \cite{papanicolaou1990}. These will have an energy gap, which we 
expect will vanish as $D \rightarrow D_c-$. For large positive D an excitation with $\Delta S^z = 2$ (i.e. $S^z = -1
\leftrightarrow S^z = +1$) 
will have a smaller energy than a single magnon. We expect, and confirm below, that there is a 2-magnon bound 
state, and we calculate its dispersion curve.

To conclude this Introduction we will briefly describe the series expansion approach, for both  ground state 
bulk properties and excitations, referring the reader to our recent book \cite{oitmaa2006} for further details. The 
approach is based on writing the Hamiltonian in the usual perturbative form

\begin{equation}
H = H_0 + \lambda V
\label{eq2}
\end{equation}
where $H_0$ has a simple known ground state. This subdivision is carried out in various ways, appropriate to 
the different phases of the model. Series are derived for the ground state energy, order parameter and other 
quantities of interest, in powers of $\lambda$, and extrapolated to $\lambda = 1$ by numerical methods 
(Pad{\' e} approximants or integrated differential approximants). Calculations are carried out for a sequence 
of finite connected clusters, and the results combined to obtain bulk lattice properties.

 A similar approach 
is used for excitations. An orthogonal transformation is used to obtain an `effective Hamiltonian' matrix for 
each cluster, yielding transition amplitudes for the excitation in real space, which are then combined to 
obtain dispersion curves throughout the Brillouin zone. Points where the gap approaches zero can be easily 
identified, and corresponding series obtained for the gap itself.

In the body of the paper we will present and analyse various results in the easy-axis ($D > 0$) phase (Section II), 
and in the easy-plane ($D < 0$) phase (Section III). Section IV contains a summary and conclusions.

\section{The Easy Axis ($D > 0$) Case}
\label{sec2}

\subsection{Bulk Ground State Properties}
\label{sec2a}

\begin{figure}
 \includegraphics[width=1.0\linewidth]{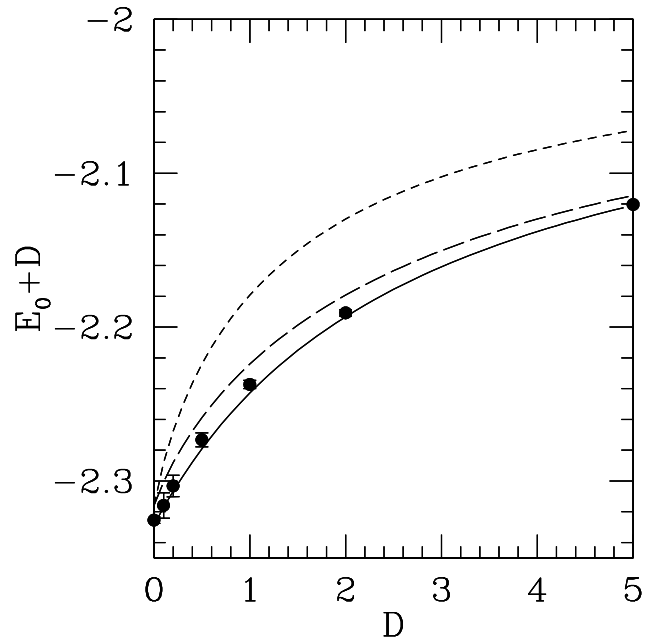} 
\caption{ Ground state energy per spin for the easy-axis J-D model on the SQ lattice. 
The individual points are the estimates from series . The various 
lines are results of different spin-wave approximations: SW1, short-dashed line; SW1a, long-dashed line; SW2,
solid line. For convenience, we plot $E_0 + D$ versus $D$, and set $J = 1$. 
 }
 \label{fig2}
\end{figure}

We have computed series for the ground state energy per spin and for the ground state staggered magnetization 
to order $\lambda^{12}$, where $H_0$ and $V$ are

\begin{eqnarray}  
H_0 & = & J \sum_{<ij>} S^z_i S^z_j -D \sum_i (S^z_i)^2 - h \sum_i \eta_i S^z_i \\
V & = & \frac{1}{2} \sum_{<ij>} (S^+_i S^-_j + S^-_i S^+_j)
\label{eq3}
\end{eqnarray}
for various values of D. Here $h$ is a staggered field, included to allow calculation of the order parameter, 
and $\eta_i = \pm 1$ on the respective sublattices.
 We do not present the series coefficients here, but can provide them on request.

\begin{figure}
 \includegraphics[width=1.0\linewidth]{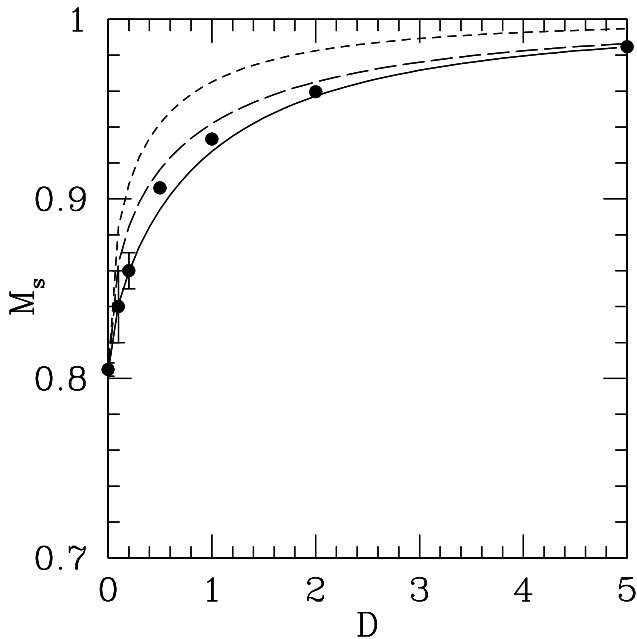}
\caption{ Staggered magnetization for the easy-axis J-D model on the SQ lattice. 
Notation as in Figure \ref{fig2}.
 }
 \label{fig3}
\end{figure}

Figures \ref{fig2} and \ref{fig3} show the ground state energy and staggered
magnetization versus D. 

For purposes of comparison, we also present results from conventional spin-wave theory, first order spin wave 
theory (SW1), modified first-order theory (SW1a) and second order theory (SW2). The first order theory gives

\begin{eqnarray}  
E_0 & = & -4J-(1+\eta)D+\frac{1}{N} \sum_k \epsilon_k \\
M_s & = & \frac{3}{2} - \frac{1}{N} \sum_k \frac{4J + 2\eta D}{\epsilon_k}
\label{eq4}
\end{eqnarray}
where $\epsilon_k$, the spin-wave energy, is given by

\begin{equation}
\epsilon_k = 4J\sqrt{(1+\eta D/(2J))^2-\gamma_k^2}
\label{eq5}
\end{equation}
with $\gamma_k = (\cos k_x + \cos k_y)/2$. The constant $\eta = 1, 1/2$ for SW1, SW1a respectively. This factor 
arises from a choice in treating the anisotropy term. In large S theory $\eta =1$ but for $S = 1$, as here, normal 
ordering of the quartic boson terms gives $\eta = 1/2$. This is explained in Appendix A, where we also present the 
more complex second-order theory.

As is apparent from Figure \ref{fig2}, SW1 is a rather poor approximation, 
but SW1a and SW2 are
in almost quantitative agreement with
the series results, within 1\%.
For the magnetization, a simlar conclusion applies.
Note that the magnetization exhibits a square-root cusp behaviour near $D = 0$, in agreement with the
spin-wave theory.

\subsection{1-Magnon Excitations}
\label{sec2b}

At least for small D the lowest energy excitations, in the unperturbed system, consist of a single spin excited 
from its ordered $S^z = \pm 1$ state to $S^z = 0$, i.e. $\Delta S^z = \pm 1$. The quantum fluctuations, embodied in the 
perturbation V, will then allow this to propagate through the lattice, forming a coherent magnon band.

Within spin wave theory the excitation energy is given by equation (\ref{eq5}) (or the more general result in 
Appendix A). We have computed a series expansion for the excitation energy to order $\lambda^{11}$, following the 
original work of Gelfand \cite{gelfand1996} (see also ref. 13).

\begin{figure}
 \includegraphics[width=1.0\linewidth]{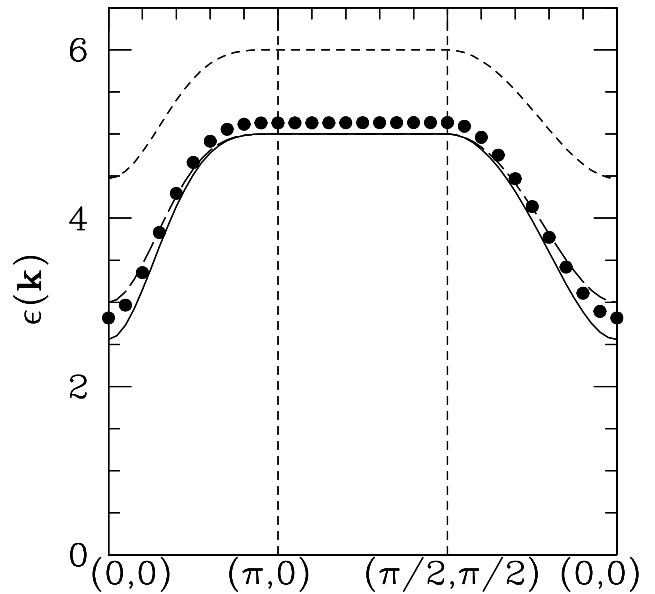}
\caption{Single-magnon excitation energy at D/J = 1.0 along symmetry lines in the Brillouin zone, obtained from series (full dots) 
and spin wave approximations. Notation as in Figure \ref{fig2}.}
 \label{fig4}
\end{figure}

In Figure \ref{fig4} we show the excitation energy, along symmetry lines in the Brillouin zone, for the case $D/J = 1$, and 
again for comparison the spin-wave results. Again we see that SW1 is a rather
poor approximation, 
but SW1a and SW2 provide an excellent description of the data, with SW1a actually better than SW2.

The dispersion curves are smooth and rather featureless. The most significant feature is the opening of a gap at 
${\bf k} = (0,0)$ for any non-zero D. This reflects, in the easy axis case, the fact that the remnant O(2) 
symmetry of the Hamiltonian is not spontaneously broken in this case, and so Goldstone modes are absent.

\begin{figure}
 \includegraphics[width=1.0\linewidth]{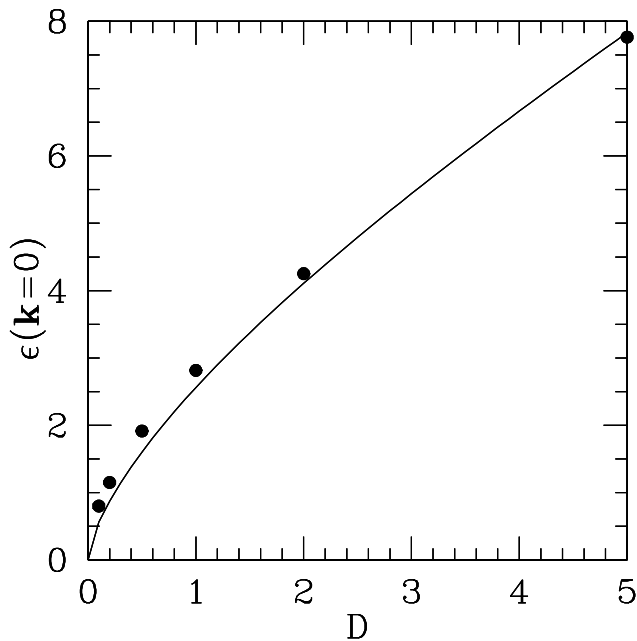}
\caption{Single-magnon excitation energy 
at ${\bf k} = (0,0)$ as a function of $D$ in the easy-axis region (setting $J = 1$). The solid line is the second order spin wave approximation (SW2).
}
 \label{fig5}
\end{figure}

Figure \ref{fig5} shows the dependence of the gap at ${\bf k} = (0,0)$ on the coupling D. The $\sqrt{D}$
dependence at small D predicted by spin wave theory is clearly evident. At large D the single-magnon gap is
predicted to increase linearly with D.

\subsection{Large D Excitations}
\label{sec2c}

It is clear that for large D the single-magnon excitations of the previous subsection will not be the lowest energy 
excitations of the system. Their energy will be of order D, whereas an excitation with $\Delta S^z =  2$ will 
have an energy of order J. Such an excitation, created at a particular site, can again propagate through the 
lattice, forming a quasiparticle band. We may think of this as a two-magnon bound state where the magnons are bound 
on the same site (see Appendix A). 

Figure \ref{fig6} shows the dispersion relation for the $\Delta S^z =
2$ excitation at $D/J = 1$ compared with the lower edge of the two-magnon continuum.
It can be seen that in the
mid-region of the plot the excitation indeed seems to lie below the
continuum, becoming a bound state at slightly below this coupling. The
energy here is close to the asymptotic limit of 8 units. In the wings of
the plot the error bars are much larger, and the excitation may not be
bound: it is possible that these facts may be related, At higher values of $D/J$, the bound state energy remains close to 8
units, so that the binding energy rises almost linearly with $D/J$.

\begin{figure}
 \includegraphics[width=1.0\linewidth]{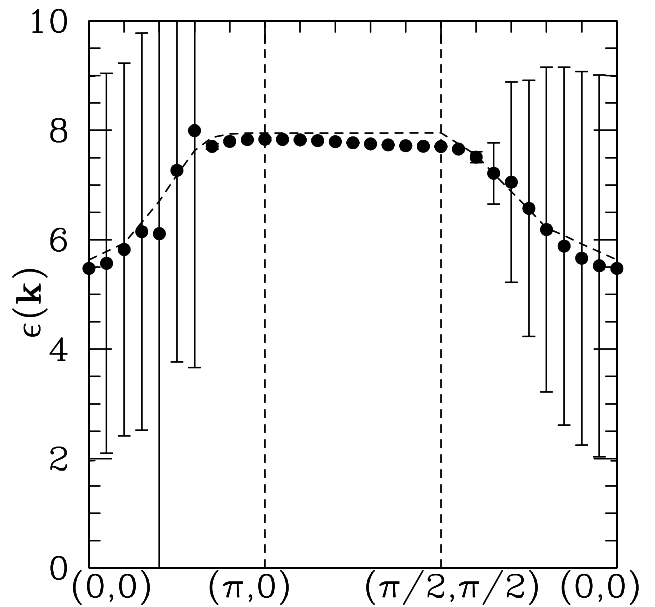}
\caption{ 
Dispersion relation of the $\Delta S^z = 2$ excitation at $D/J = 1.0$. The points with error bars are the series
estimates. The dashed line is the lower limit of the 2-magnon continuum.
}
 \label{fig6}
\end{figure}

\subsection{Finite Temperature Phase Transition}
\label{sec2d}

\begin{figure}
 \includegraphics[width=1.0\linewidth]{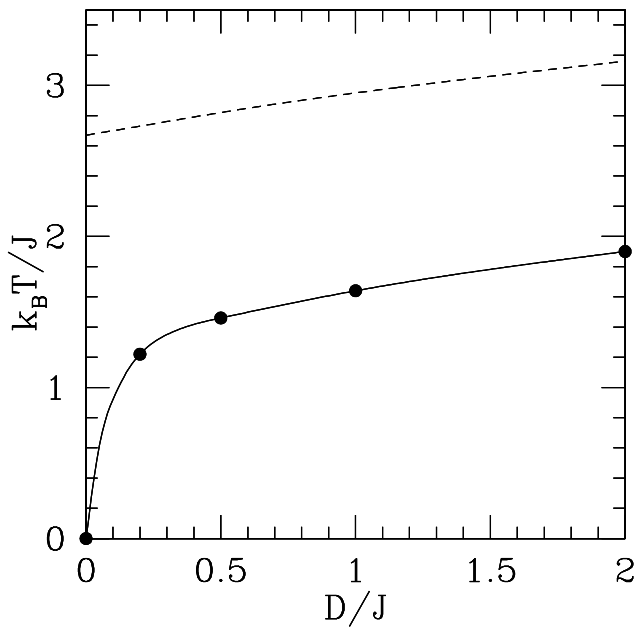}
\caption{Critical temperature versus $D/J$ for the $S = 1$ easy-axis model on the SQ lattice. The filled circles are the 
series results, with estimated errors no larger than the size of the symbols. The line is a guide to the eye. The 
dashed line is the MFA result \cite{khajehpour1975}.}
 \label{fig7}
\end{figure}

Since the continuous O(3) symmetry of the Heisenberg model is destroyed  by an easy-axis anisotropy term, 
and the order parameter is Ising-like, taking one of two possible values,
the ordered ground state will persist to finite temperature, up to a critical 
temperature $T_c(D)$. We have derived high-temperature series in the variable $K = J/k_BT$ to order $K^{11}$ for the 
staggered susceptibility $\chi_s$, for various values of D. These are then analysed via standard Dlog Pad{\' e} 
approximants to obtain the critical temperature and exponent. Figure \ref{fig7} shows the critical temperature 
versus D and, for comparison, the mean field result \cite{khajehpour1975}. One would not expect MFA to give accurate results in 
2 dimensions, and indeed there is a sizeable discrepancy. We note that MFA gives a finite critical temperature 
even for the isotropic case, $D = 0$, which violates the Mermin-Wagner theorem. The critical exponent $\gamma$ 
is consistent with the universal 2D Ising exponent 7/4, although there is substantial scatter in the estimates from 
these relatively short series.

\section{The Easy-Plane ($D < 0$) Case}
\label{sec3}

The easy-plane case shows much more interesting physics, including, as we shall see, two distinct phases separated 
by a quantum phase transition (QPT).

For small $|D|$ the spins will be preferentially in the x-y plane (choosing z as the hard axis) and the Hamiltonian 
will have O(2) symmetry. At T = 0 this symmetry will be spontaneously broken and the system will exhibit N{\' e}el 
order in some direction, reduced by quantum fluctuations. We refer to this as the planar antiferromagnetic phase (PAFM).
The broken O(2) symmetry will result in a single
gapless Goldstone mode. 
In the following we will present results from series expansions for both ground state bulk properties and for 
single-magnon excitations. These will again be compared with spin wave theory. Although there will be no 
ordered phase at finite temperature we expect a finite temperature Kosterlitz-Thouless transition. However we do not 
explore this aspect.

For large $|D|$, where the anisotropy term is dominant, we expect the system to prefer a singlet phase where spins are 
in the $S^z = 0$ state. This phase has no magnetic order, and is aptly referred to as a quantum paramagnetic phase 
(QPM). Quantum fluctuations, arising from the exchange terms, will modify the state. Low energy excitations in the 
QPM phase consist of spins excited to the $S^z = \pm 1$ states, which have been termed `excitons' and `anti-excitons'.
 In the following we derive series for both bulk properties and excitations in the QPM phase. An analytic 
approximation, due to Papanicolaou \cite{papanicolaou1990}, is also presented and compared with the series results.

As $|D|$ is reduced in the QPM phase (or increased in the PAFM phase) a quantum phase transition is found to occur and 
we use series expansions to locate this transition accurately, and to study its properties.

\subsection{PAFM Phase: Bulk Ground State Properties}
\label{sec3a}

\begin{figure}
 \includegraphics[width=1.0\linewidth]{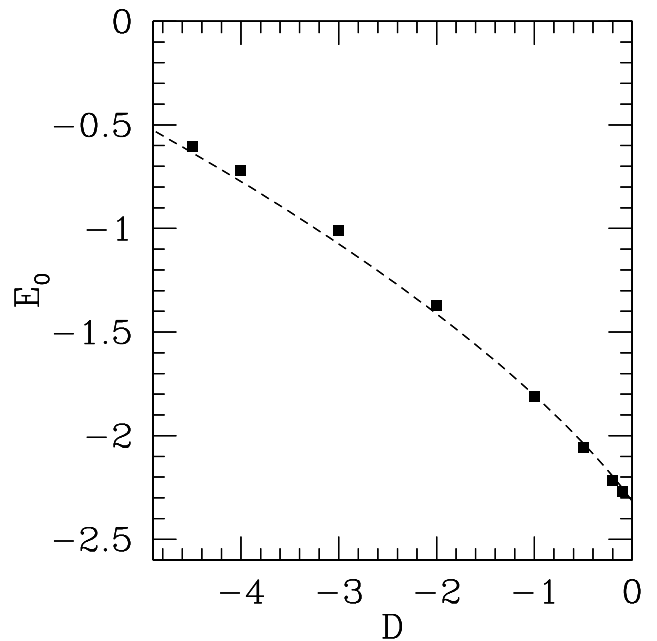} 
\caption{Ground state energy per site in the easy-plane region for the $S = 1$ J-D model on the SQ lattice. The 
squares are series results for the PAFM phase, and the 
the dashed
line is the first order spin wave estimate in the PAFM phase. We have set $J = 1$.
}
 \label{fig8}
\end{figure}

It is convenient to rotate the spin axes and to write the Hamiltonian as 
\begin{eqnarray}  
H & = & J \sum_{<ij>}(S^z_i S^z_j +S^x_i S^x_j + S^y_i S^y_j) -D \sum_i (S^x_i)^2 \nonumber \\
 & = &  J \sum_{<ij>} S^z_i S^z_j +\frac{1}{2} J \sum_{<ij>} (S^+_i S^-_j + S^-_i S^+_j) \nonumber \\
 & & -\frac{1}{4} D \sum_i 
(S^+_i +S^-_i)^2
\label{eq6}
\end{eqnarray}
where z is the ordering axis and x is the hard axis. We now decompose $H = H_0 + \lambda V$, with

\begin{eqnarray}  
H_0 & = & J \sum_{<ij>}S^z_i S^z_j +\frac{1}{2} D \sum_i (S^z_i)^2 -h \sum_i \eta_i S^z_i \\
 V & = &  \frac{1}{2} J \sum_{<ij>} (S^+_i S^-_j + S^-_i S^+_j) \nonumber \\
 & & -\frac{1}{4} D \sum_i (S^+_i S^+_i +
S^-_i S^-_i)
\label{eq7}
\end{eqnarray}
where we have dropped a constant term  $-ND$ and, as before, included a staggered field term.

\begin{figure}
 \includegraphics[width=1.0\linewidth]{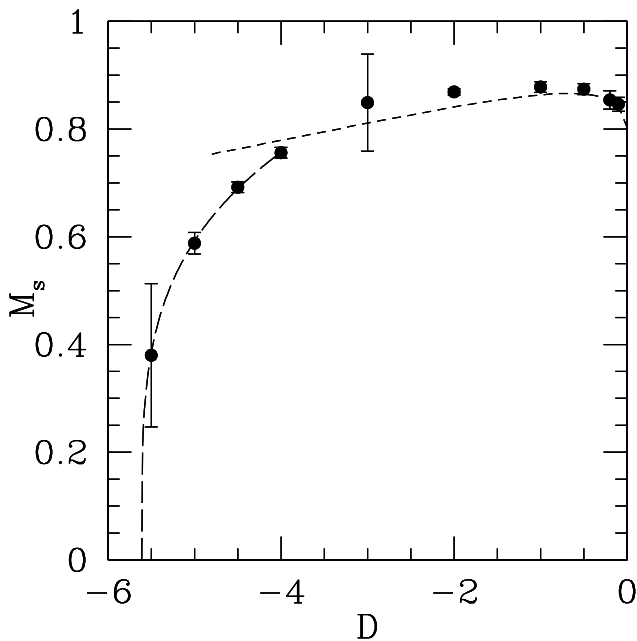} 
\caption{Staggered magnetization in the PAFM phase of the $S = 1$ J-D model on the SQ lattice. 
The points are series estimates, and the short-dashed line is first order spin wave theory.
The long-dashed line is a fit in the critical region.
}
 \label{fig9}
\end{figure}

We have computed series for the ground state energy and staggered magnetization to order $\lambda^{12}$. Figures
\ref{fig8} and \ref{fig9} show the ground state energy and magnetization versus D, as obtained by analysis of these series.

For comparison we show the results of first order spin wave theory (SW1). The details of this are given in Appendix B. 
In Figure \ref{fig8}, we see that the SW1 theory gives quite a good
description of the bulk ground state energy in the PAFM phase, as
compared with the series estimates (squares).

The series estimates for the magnetization in the PAFM phase were obtained as follows. The
perturbation series in $\lambda$ typically exhibit a singularity of form $(\lambda_c
-\lambda)^{\sigma}$ for $\lambda_c$ only a little larger than $1$, so a naive Pad{\' e} extrapolation
in $\lambda$ gives unreliable results. Instead, at each value of $D/J$ we estimate $\lambda_c$ and
$\sigma$ using standard Dlog Pad{\' e} methods, and then extrapolate the series to $\lambda = 1$ using
a variable $\delta = 1-(1-\lambda/\lambda_c)^{\sigma}$.
The data seem to show a crossover from a singularity with very small $\sigma$ at small negative $D$,
to another one with $\sigma \simeq 0.27$ nearer the critical point. The estimated magnetization at the
crossover point, around $D/J = -3.5$, shows large error bars.

The results are shown in Figure \ref{fig9}. 
We note that SW1 gives qualitatively the right behaviour at small $|D|$,
but becomes rather poor
for large $|D|$, where it shows no sign of the decrease towards the
transition point.
 Near $D = 0$, the magnetization shows a square-root cusp, the
mirror of that in the easy-axis region, marking the transition from the easy-axis to the easy-plane
phase. 
 The series results show a rapid decrease in magnetization beyond $D/J \sim -5$, heralding the 
expected quantum phase transition to the QPM phase. However the error bars are large and it is not possible to locate the transition 
with any degree of precision from the magnetization alone. The fit in this region will be discussed in
Section \ref{sec3d}.

\subsection{PAFM Phase: Elementary Excitations}
\label{sec3b}

\begin{figure}
 \includegraphics[width=1.0\linewidth]{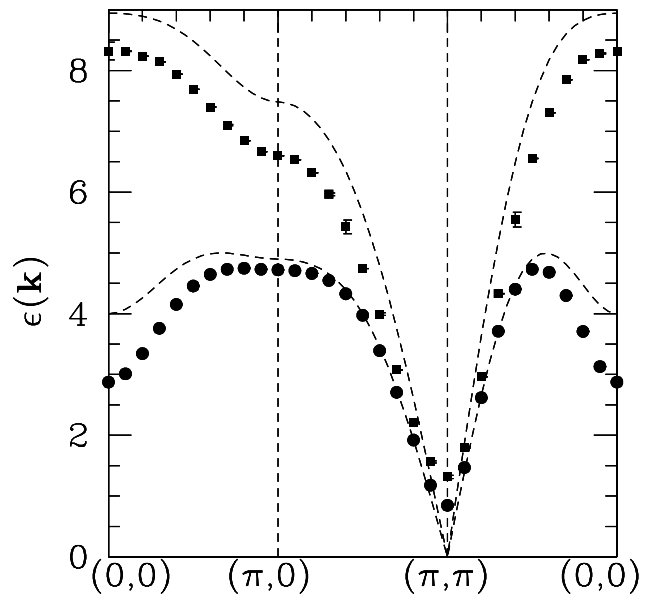} 
\caption{Single-magnon dispersion curves for the PAFM phase for $D/J = -1.0$
(circles) and $D/J = -5.0$ (squares) 
The dashed lines are from the first order spin wave theory.}
 \label{fig10}
\end{figure}

We have computed series for the single-magnon excitations in the PAFM phase, to order $\lambda^{10}$. The series are 
analysed to compute the magnon energies $\epsilon({\bf k})$, and these are 
shown in Figure \ref{fig10}, along 
symmetry lines in the Brillouin zone, for values $D/J = -1.0$ and
$-5.0$.
Again, for comparison, we show the 
result of first order spin wave theory (Appendix B). We note that spin fluctuations transverse to the ordering 
direction are no longer isotropic, and hence the full Brillouin zone must be used.

The energy gap vanishes at ${\bf k} = (\pi,\pi)$, according to spin-wave theory, corresponding to the expected Goldstone
mode. The series extrapolations in $\lambda$ by means of standard Pad{\'
e} approximants still give a finite result at 
that point, because they assume the series is regular at $\lambda = 1$.
The energy gap at ${\bf k} = (0,0)$ is indeed finite, and behaves like
$\sqrt{D}$ at small $|D|$, mirroring that in the easy-axis region. It
rises rapidly at large $D$. The qualitative behaviour is well reproduced
by SW1 theory.

\subsection{QPM Phase: Bulk Ground State Properties}
\label{sec3c}

To investigate the large $|D|$ quantum paramagnetic phase we decompose the full Hamiltonian as $H = H_0 +\lambda V$,
with
\begin{equation}
H_0 = |D| \sum_i (S^z_i)^2 + J \sum_{<ij>} S^z_i S^z_j
\label{eq8}
\end{equation}
and
\begin{equation}
 V = \frac{1}{2} \sum_{<ij>} (S^+_iS^-_j + S^-_i S^+_j)
\label{eq9}
\end{equation}

The unperturbed ground state has all spins in the $S^z = 0$ state, and the effect of the perturbation is to create (+ -)
states on neighbouring sites (exciton-antiexciton pairs). Note that, unlike the previous sections, we do not perform a
spin rotation on one sublattice. The derivation then follows standard lines, and we have obtained series to order
$\lambda^{12}$ for both the ground state energy and for the `quadrupole moment' $Q = <3(S^z_i)^2-2>$. An alternative
approach, in which only the anisotropy term is included in $H_0$ and the full exchange term as $V$, is also possible.
The expansion parameter is then $J/D$. We have carried this through but it seems not to yield any improvement in
precision.

\begin{figure}
 \includegraphics[width=1.0\linewidth]{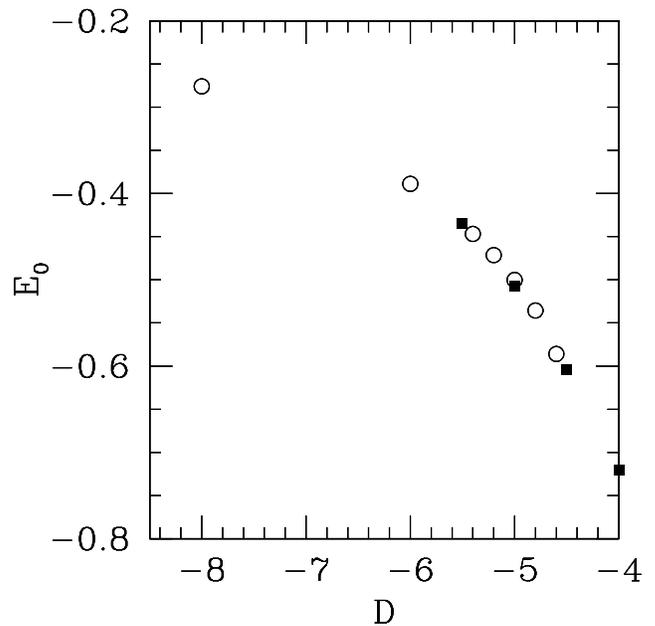}
\caption{
Ground state energy per site at negative $D/J$ (setting $J = 1$).
The open circles are estimates in the QPM phase, and the filled squares 
are PAFM estimates. 
}
 \label{fig11}
\end{figure}

In Fig. \ref{fig11} we plot the ground state energy versus $D/J$ in the QPM phase. We also include some of the data from
the PAFM expansion (Section \ref{sec3a}, above) near the crossover region. As can be seen, the two curves meet smoothly,
but the precision is inadequate to distinguish between a second order transition or a weak first order transition (with a
small discontinuity in slope). The quadrupole moment $Q$ increases smoothly from a value $-2$ at large $|D|$ to
approximately $-1.6$ at $D = -5$, but shows no interesting behaviour.

\subsection{QPM Phase: Excitations}
\label{sec3d}

\begin{figure}
 \includegraphics[width=1.0\linewidth]{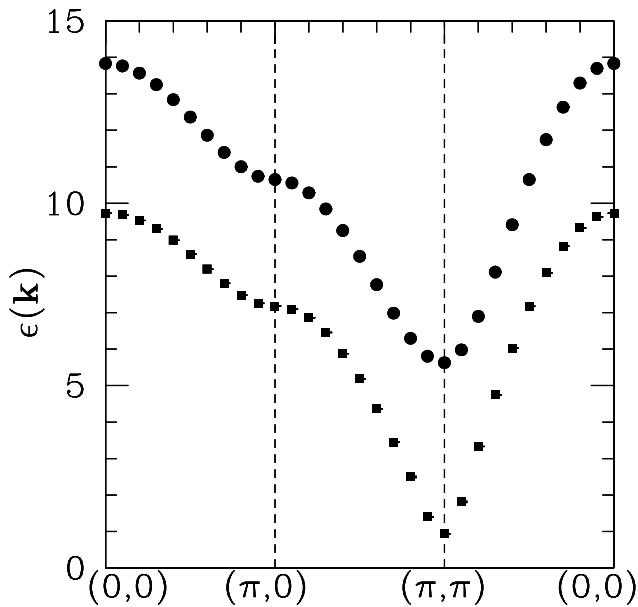}
\caption{
Single particle dispersion in the QPM phase along symmetry lines in the
Brillouin zone, for $D/J = -10.0$ (circles) and $D/J = -6.0$ (squares).
}
 \label{fig12}
\end{figure}

\begin{figure}
 \includegraphics[width=1.0\linewidth]{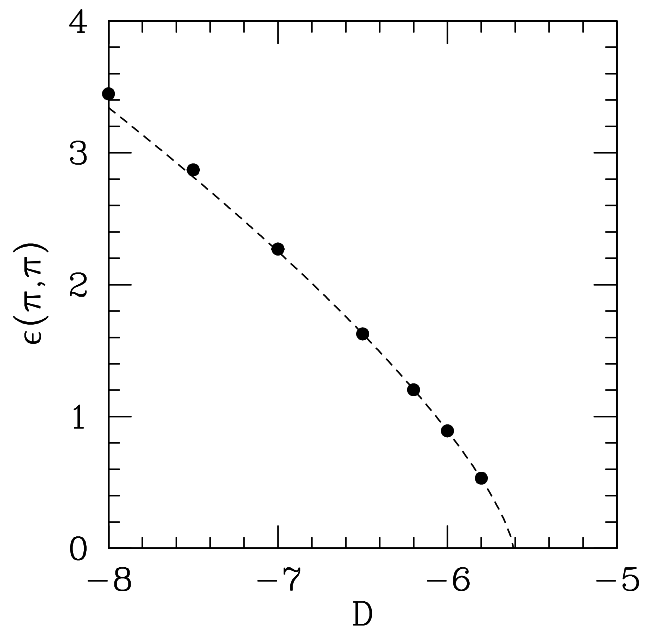}
\caption{
The energy gap $\epsilon({\bf k})$ at momentum ${\bf k} = (\pi,\pi)$ in the QPM phase, 
as a function of $D$ ($J = 1$). The dashed line
is a fit to the data in the region $-7.0 < D/J < -5.6$, in the neighbourhood of the critical point.
}
 \label{fig13}
\end{figure}

The low energy excitations, termed `excitons' and `antiexcitons', arise from exciting one of the $S^z = 0$ sites to $S^z
= \pm 1$. Such a local excitation can then propagate through the lattice as a well defined quasiparticle with energy
$\epsilon({\bf k})$.

Using the Hamiltonian decomposition (\ref{eq8},\ref{eq9}) and the usual linked cluster methods, we have computed series for the
excitation energy to order $\lambda^{10}$. Figure \ref{fig12} shows a plot along symmetry lines in the Brillouin zone
for two values of $D/J$, viz. $D/J = -10.0, -6.0$. As is apparent, the energy gap at $(\pi,\pi)$ is closing as $D$
increases, and we expect it to vanish at the quantum phase transition point $D_c$. 
We found that the best way of locating the phase transition from the
QPM to the PAFM phase was to perform a Dlog Pad{\' e} analysis of the
series in $\lambda$ for the energy gap in the QPM phase at ${\bf k} =
(\pi,\pi)$, looking for the zero point. Hence we estimate the critical
point, where the energy gap goes to zero at the physical value $\lambda
= 1$, lies at $(D/J)_c = - 5.61(5)$.  
A similar analysis of the magnetization in the PAFM phase gives a somewhat less reliable estimate, $(D/J)_c = -5.7(2)$, which is
compatible with the figure above. 
We note that the coupled cluster calculation \cite{wong1994} gives $(D/J)_c = -6.97, -6.38$ at successive levels.

The energy gap in the QPM at momentum ${\bf k} =(\pi,\pi)$ was again estimated by forming Pad{\' e}
approximants in the variable $\delta = 1 -(1-\lambda/\lambda_c)^{\sigma}$, where $\lambda_c$ and
$\sigma$ are the location and critical index, respectively, of the energy gap as a function of
$\lambda$, estimated by the usual Dlog Pad{\' e} methods. The index $\sigma$ appeared consistently as
$0.70(2)$.

Figure \ref{fig13} shows the
resulting estimates of the energy gap in the QPM at this momentum, as a function of $D/J$, with a
fit near the critical region of the form $ \epsilon({\bf k}) \propto 
(5.61-D/J)^{\nu}$, where the fit gives $\nu = 0.73(3)$.
One would naturally conclude that the critical indices $\nu$ and $\sigma$ are identical.

A similar fit of the form $M_s \propto (5.61 + D/J)^{\beta}$ to the magnetization in the PAFM phase is
shown in Figure \ref{fig9}. The fit over the range $-5.6 < D/J < -4.5$ gives $\beta = 0.25(3)$, and
again one would conclude that the magnetization indices in the variables $\lambda$ and $D/J$ are the same. 

These indices should be compared with the expected values for the universality class corresponding to
this quantum phase transition, namely those of the classical O(2) model in three dimensions, which are
$\nu = 0.671$, $\beta = 0.346$ \cite{guida1998}. The agreement is not very good, but this is perhaps
not surprising in view of the crude and indirect methods used in our estimates.

An analytic theory for the QPM phase has been proposed by Papanicolaou \cite{papanicolaou1985}, based 
on a generalized Holstein-Primakoff transformation. This gives
\begin{equation}
\epsilon({\bf k}) = \sqrt{D(D+8J\gamma_{{\bf k}})}
\label{eq10}
\end{equation}
which yields for the $(\pi,\pi)$ gap
\begin{equation}
\epsilon(\pi,\pi) = \sqrt{D(D+8J)}
\label{eq11}
\end{equation}
i.e. $(D/J)_c = -8.0$, $\nu = 0.5$. These do not agree well with the series estimates.

\section{Conclusions}
\label{Sec4}

Magnetic systems with $S = 1$ and with easy-axis or easy-plane crystal
field anisotropy have become of interest again, as a result of new
materials and suggestions of novel quantum phases. Early theoretical
approaches, based on mean field, Green's function, and spin-wave
approximations are of uncertain and dubious validity. Present day
(numerical) approaches such as Quantum Monte Carlo and series methods
allow rather precise calculation of ground state properties and of the
spectrum of elementary excitations, and hence of energy gaps.

We have used comprehensive series methods, for the first time, to study
the model on the square lattice. Three distinct phases are identified,
in agreement with previous work. We have compared our results with those
of first and second order spin-wave theory. In the Ising
antiferromagnetic (IAFM) phase, the first order theory deviates
substantially from the series results, but the second order theory (and
even a 'modified' first order theory) are in quantitative agreement with
series, within one or two percent. In the planar antiferromagnetic
(PAFM) phase, only a first order theory is available, which gives a
reasonable description at small negative $D$, but fails in the
neighbourhood of the transition to the quantum paramagnetic (QPM) phase. 

The transition between the planar
antiferromagnetic and quantum paramagnetic phase is located
at $D/J = -5.61(5)$. The transition appears to be of second order, with
critical indices in qualitative but not quantitative agreement with
those of the classical O(2) model in three spatial dimensions. We would
not claim any contradiction here as the error bars on our estimates are
rather large.

The series approach followed in this paper can, of course, be applied
equally well to other lattices. Indeed there is considerable interest in
the one-dimensional case, and work on this is in progress.

\acknowledgments
This work forms part of a research project supported by a grant
from the Australian Research Council.
We are grateful for the
computing resources provided by the Australian Partnership for
Advanced Computing (APAC) National Facility and by the Australian
Centre for Advanced Computing and Communications (AC3).

\begin{widetext}
\appendix
\section{{\bf Spin Wave Theory for the Easy-Axis ($D > 0$) Case}}
\label{appA}

The spin-wave approximation is well known. Nevertheless, for completeness, we give a brief summary of the second order 
theory for this model, following closely the treatment of ref. \cite{zheng1991}.

The initial Hamiltonian

\begin{equation}
H = J \sum_{<ij>} [ S^z_i S^z_j +\frac{1}{2}(S^+_i S^-_j + S^-_i S^+_j)] -D \sum_i (S^z_i)^2
\label{eqA1}
\end{equation}
is expressed in terms of boson operators $a_i, b_j$ on the respective sublattices via a Dyson-Maleev 
transformation
\begin{eqnarray}
\begin{array}{ccc}
A: \hspace{5mm}S^z_i = S - a^{\dagger}_i a_i,\hspace{5mm} & S^+_i = \sqrt{2S}(1-a^{\dagger}_i a_i/2S)a_i, \hspace{5mm}& 
S^-_i = \sqrt{2S} a^{\dagger}_i \\
B: \hspace{5mm} S^z_j = b^{\dagger}_j b_j -S, \hspace{5mm}& S^+_j = \sqrt{2S}b^{\dagger}_j(1-b^{\dagger}_j b_j/{2S}),
\hspace{5mm} 
& S^-_j = \sqrt{2S} b_j
\end{array}
\label{eqA2}
\end{eqnarray}
followed by a transformation to {\bf k}-space (Bloch) operators
\begin{equation}
a_i = \sqrt{\frac{2}{N}} \sum_{{\bf k}} e^{-i{\bf k \cdot R_i}} a_{\bf k}, \hspace{5mm} 
b_j = \sqrt{\frac{2}{N}} \sum_{{\bf k}} e^{i{\bf k.R_j}} b_{{\bf k}}
\label{eqa3}
\end{equation}
where the sum is over N/2 points in the reduced Brillouin zone, giving
\begin{eqnarray}
H & = & -\frac{1}{2} N S^2 (zJ+2D) +[S(zJ+2D)-D] \sum_{\bf k} (a^{\dagger}_{\bf k}a_{\bf k} +
b^{\dagger}_{\bf k} b_{\bf k}) + zJS \sum_{\bf k} \gamma_{\bf k} (a^{\dagger}_{\bf k} 
b^{\dagger}_{\bf k} +a_{\bf k}b_{\bf k}) \nonumber \\
 & & -\frac{zJ}{N} \sum_{{\bf k_1,k_2,k_3,k_4}} \Delta({\bf k_1-k_2-k_3+k_4}) [ 2\gamma_{{\bf k_3-k_4}}
a^{\dagger}_{{\bf k_1}} a_{{\bf k_2}}b^{\dagger}_{{\bf k_3}}b_{{\bf k_4}} +
\gamma_{{\bf k_4}} a^{\dagger}_{{\bf k_1}} a_{{\bf k_2}} a_{{\bf k_3}} b_{{\bf k_4}}+
\gamma_{{\bf k_1}} a^{\dagger}_{{\bf k_1}}b^{\dagger}_{{\bf k_2}} b^{\dagger}_{{\bf k_3}}b_{{\bf k_4}}] 
\nonumber \\
 & & -\frac{2D}{N} \sum_{{\bf k_1,k_2,k_3,k_4}}\delta({{\bf k_1+k_2-k_3-k_4}}) [a^{\dagger}_{{\bf k_1}}a^{\dagger}_{{\bf k_2}}
a_{{\bf k_3}}a_{{\bf k_4}}+b^{\dagger}_{{\bf k_1}}b^{\dagger}_{{\bf k_2}}b_{{\bf k_3}}b_{{\bf k_4}}]
\label{eqa4}
\end{eqnarray}
Here $z$ is the coordination number of the lattice and $\gamma_{{\bf k}}$ is the usual
\begin{equation} 
\gamma_{{\bf k}} = 1/z \sum_{nn} \exp{i{\bf k \cdot \delta}} = \frac{1}{2}(\cos k_x + \cos k_y) 
\end{equation} 
for the SQ lattice. 

The reader's attention is drawn to the factor $[S(zJ+2D)-D]$ associated with the diagonal quadratic terms. If we 
consider successive orders of spin wave theory as corresponding to decreasing powers of $S$, then the first-order 
theory (SW1) will retain only $S(zJ+2D)$. However if we include the complete term, for $S=1$, we have $(zJ+D)$. This is 
the origin of the modified first-order theory (SW1a) discussed in Section \ref{sec2a}.

To diagonalize the quadratic part of the Hamiltonian we use a standard Bogoliubov transformation
\begin{eqnarray}
a_{{\bf k}} & = & u_{{\bf k}}A_{{\bf k}} - v_{{\bf k}}B^{\dagger}_{{\bf k}} \nonumber \\
b_{{\bf k}} & = & -v_{{\bf k}}A^{\dagger}_{{\bf k}} +u_{{\bf k}}B_{{\bf k}} 
\label{eqA5}
\end{eqnarray}
with $u_{{\bf k}} = \cosh \theta_{{\bf k}}, v_{{\bf k}} = \sinh \theta_{{\bf k}}$.

This gives, after some algebra,
\begin{eqnarray}
H & = & NE_0 +\sum_{{\bf k}} \Omega_{{\bf k}}(A^{\dagger}_{{\bf k}}A_{{\bf k}} + 
B^{\dagger}_{{\bf k}}B_{{\bf k}}) + \sum_{{\bf k}} V_{{\bf k}}(A^{\dagger}_{{\bf k}}B^{\dagger}_{{\bf k}} + 
A_{{\bf k}}B_{{\bf k}})  \nonumber \\
 & &  
+\frac{1}{N}\sum_{{\bf k_1,k_2,k_3,k_4}} \delta({{\bf k_1+k_2-k_3-k_4}}) V_4({{\bf k_1,k_2,k_3,k_4}}) (B^{\dagger}_{{\bf
k_1}}B^{\dagger}_{{\bf k_2}}B_{{\bf k_3}}B_{{\bf k_4}} + A^{\dagger}_{{\bf k_1}}A^{\dagger}_{{\bf k_2}}A_{{\bf k_3}}A_{{\bf
k_4}}) \nonumber \\
 & & +
further \ normal  \ ordered  \ quartic  \ terms
\label{eqA6}
\end{eqnarray}
where
\begin{eqnarray}
E_0 & = & -(2J+D)+(4J+D)R_2-4JR_3-2J(R_2-R_3)^2-2DR_2^2 \\
\Omega_{{\bf k}} & = & [4J(1-R_2+R_3)+D(1-4R_2)]\cosh 2\theta_{{\bf k}} -4J(1-R_2+R_3)
\gamma_{{\bf k}}\sinh 2\theta_{{\bf k}} \\
V_{{\bf k}} & = & 4J(1-R_2+R_3)\gamma_{{\bf k}} \cosh 2 \theta_{{\bf k}} -[4J(1-R_2+R_3)
+D(1-4R_2)] \sinh 2\theta_{{\bf k}}  \\
V_4({\bf k_1,k_2,k_3,k_4}) & = & 4J[(u_{{\bf k_1}}u_{{\bf k_2}}u_{{\bf k_3}}v_{{\bf k_4}} + v_{{\bf k_1}}v_{{\bf k_2}}v_{{\bf
k_3}}u_{{\bf k_4}})\gamma_{{\bf k_4}}-2u_{{\bf k_1}}v_{{\bf k_2}}u_{{\bf k_3}}v_{{\bf k_4}}\gamma_{{\bf k_2-k_4}}]
\nonumber \\
 & & -2D[u_{{\bf k_1}}u_{{\bf k_2}}u_{{\bf k_3}}u_{{\bf k_4}}+v_{{\bf k_1}}v_{{\bf k_2}}v_{{\bf k_3}}v_{{\bf k_4}}] 
\label{eqA7}
\end{eqnarray}
and
\begin{eqnarray}
R_2 & = & 
\frac{2}{N} \sum_{{\bf k}} v_{{\bf k}}^2  =  -\frac{1}{2} + \frac{1}{N} \sum_{{\bf k}} \cosh 2\theta_{{\bf k}} \nonumber
\\
R_3 & = & \frac{2}{N} \sum_{{\bf k}} \gamma_{{\bf k}}u_{{\bf k}}v_{{\bf k}} = \frac{1}{N} \sum_{{\bf k}} 
\gamma_{{\bf k}} \sinh 2\theta_{{\bf k}}
\label{eqA8}
\end{eqnarray}
and we have set $z=4, S=1$.

We may choose our parameter $\theta_{{\bf k}}$ so that $V_{{\bf k}} = 0$, i.e.
\begin{equation}
\tanh 2\theta_{{\bf k}} = \frac{4J(1-R_2+R_3)\gamma_{{\bf k}}}{4J(1-R_2+R_3)+D(1-4R_2)}
\label{eqa9}
\end{equation}
Dropping the quartic terms in (\ref{eqA6}) then yields the second order spin-wave Hamiltonian
\begin{equation}
H = NE_0 + \sum_{{\bf k}} \epsilon_{{\bf k}} (A^{\dagger}_{{\bf k}}A_{{\bf k}} + B^{\dagger}_{{\bf k}}B_{{\bf k}})
\label{eqA10}
\end{equation}
with
\begin{equation}
\epsilon_{{\bf k}}^2 = [4J(1-R_2+R_3)+D(1-4R_2)]^2 -[4J(1-R_2+R_3)\gamma_{{\bf k}}]^2
\label{eqA11}
\end{equation}

The magnetization is 
\begin{equation}
M = S - <a^{\dagger}_ia_i> =  \cdots = 1-R_2
\label{eqA12}
\end{equation}

These equations can then be solved numerically. Note that the expressions (\ref{eqA8}) for $R_2$ and $R_3$ 
themselves involve $R_2$ and $R_3$ on the right-hand side, and must be solved iteratively. A convenient starting 
point is the first-order spin-wave results. We used a double Gaussian quadrature procedure to carry out the Brillouin 
zone integrations.

We can obtain the qualitative behaviour of these quantities at small $D$ from the SW1a approximation. Then 
\begin{eqnarray}
\epsilon_{\bf k} & = & \sqrt{(4J+D)^2-(4J\gamma_{\bf k})^2} \nonumber \\
 & \sim & 2\sqrt{2JD} \hspace{5mm} as \hspace{5mm} D \rightarrow 0
\label{eqa13}
\end{eqnarray}
for ${\bf k} = (0,0)$, showing that the energy gap behaves like $\sqrt{D}$ at small $D$. In the same approximation, we
find
\begin{eqnarray}
R_2 & = & \frac{1}{8\pi^2} \int_0^{2\pi} \int_0^{2\pi} dk_x dk_y \frac{(1+D/4J)}{[(1+D/4J)^2-(\cos k_x + \cos k_y)^2/4]}
-\frac{1}{2} \nonumber \\
 & \sim & 0.1966 -\frac{1}{\pi}\left( \frac{D}{2J} \right)^{1/2} as \hspace{5mm} D \rightarrow 0
\label{eqa14}
\end{eqnarray}
using the results of \cite{zheng1991}. Thus we see a $\sqrt{D}$ singularity emerging in the magnetization near $D
\rightarrow 0$ as well.

At large $D$, on the other hand, we have 
\begin{equation}
u_{{\bf k}} \sim 1, \hspace{5mm} v_{{\bf k}} \sim 0 \hspace{5mm} as \hspace{5mm}D \rightarrow\infty 
\label{eqa15}
\end{equation}
and hence in leading order the single-magnon energy is
\begin{equation}
\epsilon_{{\bf k}} \sim D+4J \hspace{5mm} as \hspace{5mm}D \rightarrow \infty
\label{eqa16}
\end{equation}
The 2-particle transition amplitude is
\begin{equation}
V_4({{\bf k_1,k_2,k_3,k_4}}) \sim -2D \hspace{5mm} as \hspace{5mm} D \rightarrow \infty
\label{eqa17}
\end{equation}
In terms of the centre-of-mass and relative momenta
\begin{equation}
{\bf K} = {\bf k_1 + k_2} = {\bf k_3+k_4}
\label{eqa18}
\end{equation}
\begin{equation}
{\bf q} = \frac{1}{2}({\bf k_1 - k_2}), \hspace{5mm} {\bf p} = \frac{1}{2}({\bf k_3-k_4})
\label{eqa19}
\end{equation}
the 2-particle bound state obeys the integral Bethe-Salpeter equation
\begin{equation}
[E_2({\bf K}) - E_1({\bf K/2+q})-E_1({\bf K/2-q})]\psi_2({\bf K,q}) = \frac{1}{N}\sum_{{\bf p}} M({\bf K,q,p}) \psi_2({\bf
K,p})
\label{eqa20}
\end{equation}
where
\begin{equation}
M({\bf K,q,p}) = 2V_4({\bf K,q,p}) \sim -4D \hspace{5mm} as \hspace{5mm}
D \rightarrow \infty.
\label{eqa21}
\end{equation}
This equation is satisfied by a solution where $\psi_2({\bf K,q})$ is {\it independent} of ${\bf q}$ (corresponding to
two particle excitations at the same point), with
\begin{eqnarray}
E_2({\bf K}) & = &E_1({\bf K/2+q})+E_1({\bf K/2-q}) -2D \nonumber \\
 & \sim & 2(4J+D)-2D = 8J \hspace{5mm} as \hspace{5mm}D \rightarrow \infty
\label{eqa22}
\end{eqnarray}
This is precisely the energy one would naively expect for a $\Delta S^z = \pm 2$ excitation in
this limit.

\section{{\bf Spin Wave Theory for the PAFM Phase}}
\label{appB}

To derive spin-wave theories for the easy-plane small $|D|$ phase, we assume 2-sublattice N{\' e}el order in the 
z-direction and write the Hamiltonian as 
\begin{equation}
H = J \sum_{<ij>} [S^z_iS^z_j +\frac{1}{2} (S^+_iS^-_j + S^-_i S^+_j)] -\frac{1}{4}D \sum_i (S^+_i+S^-_i)^2
\label{eqB1}
\end{equation}
where we have chosen the x-axis to be the hard direction.

To leading order in S we again introduce boson operators in the respective sublattices
\begin{eqnarray}
\begin{array}{ccc}
A: \hspace{5mm}S^z_i = S-a^{\dagger}_ia_i, \hspace{5mm}& S^+_i = \sqrt{2S}a_i, \hspace{5mm}& S^-_i = \sqrt{2S}a^{\dagger}_i \\
B: \hspace{5mm}S_j = b^{\dagger}_jb_j - S, \hspace{5mm}& S^+_j = \sqrt{2S}b^{\dagger}_j, \hspace{5mm}& S^-_j = \sqrt{2S}b_j
\end{array}
\end{eqnarray}
followed by a transformation to Bloch operators as before. Keeping only quadratic terms yields
\begin{eqnarray}
H & = & -(2J+\frac{1}{2}D)N +(4J-D) \sum_{{\bf k}} (a^{\dagger}_{{\bf k}}a_{{\bf k}}+
b^{\dagger}_{{\bf k}}b_{{\bf k}}) +4J \sum_{{\bf k}}\gamma_{{\bf k}}(a^{\dagger}_{{\bf k}}
b^{\dagger}_{{\bf k}}+a_{{\bf k}}b_{{\bf k}} \nonumber \\ 
 & & -\frac{1}{2} D \sum_{{\bf k}} (a^{\dagger}_{{\bf k}}a^{\dagger}_{{\bf -k}} +a_{{\bf k}}a_{{\bf -k}} +
b^{\dagger}_{{\bf k}}b^{\dagger}_{{\bf -k}} +b_{{\bf k}}b_{{\bf -k}}
\label{eqB2}
\end{eqnarray}
where we have set $S=1, z=4$. The Brillouin zone sums are over $N/2$ points in the reduced zone.

To diagonalize this, more general, quadratic Hamiltonian we introduce 
operators $\{q_1,q_2,q_3,q_4\} \equiv \{a_{{\bf k}}, 
b^{\dagger}_{{\bf k}},a^{\dagger}_{{\bf -k}},b_{{\bf -k}}\}$ and write the Hamiltonian as 
\begin{equation}
H = -4JN +\frac{1}{2}\sum_{{\bf k}}h_{ij}q^{\dagger}_iq_j
\label{eqB3}
\end{equation}
where ${\bf h}$ is the 4 x 4 matrix
\begin{eqnarray}
{\bf h} = \left( \begin{array}{cccc}
4J-D & 4J\gamma_{{\bf k}} & -D & 0 \\
4J\gamma_{{\bf k}} & 4J-D & 0 & -D \\
-D & 0 & 4J-D & 4J\gamma_{{\bf k}} \\
0 & -D & 4J\gamma_{{\bf k}} & 4J-D
\end{array} \right)
\label{eqB4}
\end{eqnarray}

We now introduce a transformation to new coordinates $\{Q_i: i=1,4\}$
\begin{equation}
q_i = \sum_j S_{ij}Q_j
\label{eqB5}
\end{equation}
with the constraint $[Q_i, Q^{\dagger}_j] = J_i \delta_{ij}$ with 
$J_i =(1,-1,-1,1)$, i.e. $SJS^{\dagger}=J$ 
 (where ${\bf J}$ is a diagonal matrix with entries $J_{ii} = J_i$). 
Following
the argument of Tsallis \cite{tsallis1978}, one easily shows that the matrix 
\begin{eqnarray}
{\bf \tilde{h}} = {\bf hJ} = \left( \begin{array}{cccc}
4J-D & -4J\gamma_{{\bf k}} & D & 0 \\
4J\gamma_{{\bf k}} & -(4J-D) & 0 & -D  \\
-D & 0 & -(4J-D) & 4J \gamma_{{\bf k}} \\
0 & D & -4J\gamma_{{\bf k}} & 4J-D
\end{array} \right)
\label{eqB5}
\end{eqnarray}
can be diagonalized by a similarity transformation,
with its eigenvalues remaining invariant. 
It has eigenvalues $\pm \lambda_1,\pm \lambda_2$ where $\lambda_1,\lambda_2$ are the spin-wave energies.

The diagonalized Hamiltonian can then be written as
\begin{equation}
H = NE_0 +\sum_{{\bf k}} [\lambda_{1{\bf k}}A^{\dagger}_{{\bf k}}A_{{\bf k}} + 
\lambda_{2{\bf k}}B^{\dagger}_{{\bf k}}B_{{\bf k}}]
\label{eqB6}
\end{equation}
with
\begin{equation}
E_0 = -4J +\frac{1}{2N} \sum_{{\bf k}}(\lambda_{1{\bf k}}+\lambda_{2{\bf k}})
\label{eqB7}
\end{equation}
Direct calculation gives
\begin{eqnarray}
\lambda_{1{\bf k}}^2 & = & 16J^2(1-\gamma_{{\bf k}}^2) - 8DJ(1+\gamma_{{\bf k}}) \\
\lambda_{2{\bf k}}^2 & = & 16J^2(1-\gamma_{{\bf k}}^2) - 8DJ(1-\gamma_{{\bf k}}) 
\label{eqB8}
\end{eqnarray}
In the reduced zone we have two branches, one of which is gapless at ${\bf k} = (0,0)$ and the other at $(\pi,\pi)$.
However we note that $\lambda_2(\pi-k_x,\pi-k_y) = \lambda_1(k_x,k_y)$ and hence in a full zone we need only consider a 
single branch $\omega_{{\bf k}} = \lambda_{1{\bf k}}$. Then we find that the spin wave energy vanishes at ${\bf k} =
(\pi,\pi)$, corresponding to the expected Goldstone mode, while at ${\bf k} = (0,0)$ the gap is $4\sqrt{J(-D)}$, mirroring
the square root behaviour found in the easy-axis case (modulo the factor $\eta$ referred to previously).

The magnetization is given by
\begin{equation}
M = 1-\frac{2}{N} \sum_{{\bf k}} [|S_{12}({\bf k})|^2 + |S_{13}({\bf k})|^2]
\label{eqB9}
\end{equation}
and can be obtained numerically from the transformation equations.

The theory described above follows from either the Holstein-Primakoff or Dyson-Maleev approach, at lowest order. 
However an attempt to extend these to higher order fails, as the resulting spin wave energies do not possess the 
Goldstone mode required by symmetry.

\end{widetext}


\begin{references}
\bibitem{birgeneau1970} R. J. Birgeneau, J. Skalyo, Jr., and G. Shirane, J. Appl. Phys. {\bf 41}, 1303 (1970).
\bibitem{steiner1987} M. Steiner et al., J. Appl. Phys. {\bf 61}, 3953 (1987).
\bibitem{dorner1988} B. Dorner et al., Z. Phys. B{\bf 72}, 487 (1988).
\bibitem{renard1988} J.P. Renard et al., J. Appl. Phys. {\bf 63}, 3538 (1988).
\bibitem{orendac1995} M. Orendac et al., Phys. Rev. B{\bf 52}, 3435 (1995).
\bibitem{haldane1983} F.D.M. Haldane, Phys. Rev. Lett. {\bf 50}, 1153 (1983).
\bibitem{murakami1996} Y. Murakami and H. Suematsu, Phys. Rev. B{\bf 54}, 4146 (1996).
\bibitem{uchida2002} M. Uchida {\it et al.}, Phys. Rev. B{\bf 66},
054429 (2002).
\bibitem{stone2008} M.B. Stone {\it et al.}, cond-mat.str-el/08012332v1.
\bibitem{nakatsuji2005} S. Nakatsuji et al., Science {\bf 309}, 1698 (2005).
\bibitem{tsunetsugu2006} H. Tsunetsugu and M. Arikawa, J. Phys. Soc. Japan {\bf 75}, 083701 (2006).
\bibitem{bhattacharjee2006} S. Bhattacharjee, V.B. Shenoy and T. Senthil, Phys. Rev. B{\bf 74}, 092406 (2006).
\bibitem{chandra1991} P. Chandra and P. Coleman, Phys. Rev. Lett. {\bf 66}, 100 (1991).
\bibitem{greiner2002} M. Greiner et al., Nature {\bf 415}, 39 (2002).
\bibitem{oitmaa2006} J. Oitmaa, C.J. Hamer and W. Zheng, {\it `Series Expansion Methods for Strongly Interacting Lattice Models'} (Cambridge University Press, 2006).
\bibitem{sachdev1999} S. Sachdev, {\it `Quantum Phase Transitions'} (Cambridge University Press, 1999).
\bibitem{golinelli1992} O. Golinelli, Th. Jolicoeur and R. Lacaze, Phys. Rev. B{\bf 45}, 9798 (1992); and references therein.
\bibitem{papanicolaou1990} N. Papanicolaou and P. Spathis, J. Phys. Cond. Mat. {\bf 2}, 6575 (1990).
\bibitem{chen2003} W. Chen, K. Hida and B.C. Sanctuary, Phys. Rev. B{\bf 67}, 104401 (2003).
\bibitem{devlin1971} J. Devlin, Phys. Rev. B{\bf 4}, 136 (1971).
\bibitem{khajehpour1975} M.R.H. Khajepour et al., Phys. Rev. B{\bf 12}, 1849 (1975).
\bibitem{rastelli1974} E. Rastelli et al., J. Phys. C: Solid State {\bf 7}, 1735 (1974).
\bibitem{lindgard1975} P.A. Lindgard et al., J. Phys. C: Solid State {\bf 8}, 1059 (1975).
\bibitem{wong1994} W.H. Wong et al., Phys. Rev. B{\bf 50}, 6126 (1994).
\bibitem{neves1986} E.J. Neves and J.F. Perez, Phys. Lett. A{\bf 114},
331 (1986).
\bibitem{singh1990} R.R.P. Singh, Phys. Rev. B{\bf 41}, 4873 (1990).
\bibitem{gelfand1996} M.P. Gelfand, Solid State Comm. {\bf 98}, 11 (1996).
\bibitem{guida1998} R. Guida and J. Zinn-Justin, J. Phys. A: Math Gen {\bf 31}, 8103 (1998).
\bibitem{papanicolaou1985} N. Papanicolaou, Z. Phys. B{\bf 61}, 159 (1985).
\bibitem{zheng1991} W-H. Zheng, J. Oitmaa and C.J. Hamer, Phys. Rev.
B{\bf 43}, 8321 (1991).
\bibitem{tsallis1978} C. Tsallis, J. Math. Phys. {\bf 19}, 277 (1978).
\end{references}
\end{document}